# TIME PROFILE OF U.S. NEIGHBORHOODS: DATASETS OF TIME USE AT SOCIAL INFRASTRUCTURE PLACES

**Yan Wang**
Department of Urban and Regional Planning and Florida Institute for Built Environment,
University of Florida,
P.O. Box 115706, Gainesville, FL 32611.
yanw@ufl.edu

**Ziyi Guo**
Department of Urban and Regional Planning and Florida Institute for Built Environment Resilience,
University of Florida,
Gainesville, FL 32611.

## ABSTRACT

Social infrastructure plays a critical role in shaping neighborhood well-being by fostering social and cultural interaction, enabling service provision, and encouraging exposure to diverse environments. Despite the growing knowledge of its spatial accessibility, time use at social-infrastructure places is underexplored due to the lack of a spatially resolved national dataset. We address this gap by developing scalable Social-infrastructure Time Use measures (STU) that capture length and depth of engagement, activity diversity, and spatial inequality, supported by the first-of-their-kind datasets spanning multiple geographic scales—from census tracts to metropolitan areas. Our datasets leverage anonymized and aggregated foot traffic data collected between 2019 and 2024 across 49 continental U.S. states. The data description reveals variances in STU across time, space, and differing neighborhood socio-demographic characteristics. Validation demonstrates generally robust population representation, consistent with established national survey findings while revealing more nuanced patterns. Future analyses could link STU with public health outcomes and environmental factors to inform targeted interventions aimed at enhancing population well-being and guiding social infrastructure planning and usage.

**K**eywords Activity Diversity, Mobility Data, Social Infrastructure, Time Poverty, Time Use

## 1 Background & Summary

Understanding time use offers a unique lens into population well-being, economics, social equity, and even poverty as it reveals how individuals allocate their most finite resource – time – across various activities[1]. Among essential activities such as work, education, personal care and leisure, extant urban and regional studies have emphasized the importance of social interactions and the role of urban form for urban vibrancy[2,3]. Especially, time spent at social infrastructure places, such as community centers, churches, meal sites, and cultural markets, emerges as a vital indicator of population-level social interactions[4–6], as these places foster social capital and meaningful interactions within a community[7–9]. Despite the extensive use of national datasets like the American Time Use Survey (ATUS) to explore demographic variations in time use – including factors like gender, age, marital status, parenthood, health status, and educational attainment among others[10–12] – there is currently no spatially-resolved national dataset that integrates time use at social infrastructure places across space, hindering advancements in both time geography and public health, among other interdisciplinary research.

The concept of time geography has been instrumental in studying the allocation of time across activities and spaces, providing insights into human contacts and accessibility[13,14]. The specific area of travel-time accessibility has focused on measuring and mapping travel time to characterize accessibility across large geographical areas, with growing efforts on analyzing healthcare facilities[4,15,16]. Fewer studies have been found to map and analyze the dwell time across facilities and places. Interdisciplinary studies of built environment and health have also revealed that the well-being of neighborhood residents is related to the daily activities that they engage in and the geospatial contexts that shape those social behaviors [17–19]. Given the health and well-being implications of time use[18,19], researchers in the built environment disciplines have investigated the interplay between urban form and time use, primarily through self-reported surveys across various geographical scales and in siloed cases [3,20,21]. These studies often yield mixed results, partly due to conscious or unconscious respondent biases[22]. The lack of generalizable and scalable findings underscores the need for more comprehensive, objective, large-scale time use datasets to elucidate relationships between urban spatial forms and time use patterns[3,20,23].

We initiate the effort to develop four new measures of neighborhood Social infrastructure Time Use (STU) and their datasets across the 49 continental states of the United States. We expect the new scientific data to be used in scalable empirical studies to bridge the identified gaps in time use literature across disciplines. The novel STU measures are developed around three key dimensions of neighbourhoods: the length and depth of engagement, activity diversity, and spatial inequality. Our STU datasets of the measures provide a week-by-week time series documentation from January 2019 to May 2024 across spatial scales, from census tract to metropolitan area.

We use census tracts – which typically have a population size ranging from 1,200 to 8,000 people – as proxies for neighborhoods, due to their widespread acceptance as a reliable analytical spatial unit. Such spatial unit also enables linkages with existing public health and environmental datasets such as CDC's PLACES [24] and the Climate Vulnerability Index [25]. The neighborhood-level data will allow detailed examinations of both compositional (varying populations) and contextual (differing places) factors influencing health outcomes [26], facilitating the formulation of targeted neighborhood-level interventions designed to promote health equity [27]. We have further aggregated these neighborhood measurements and data into larger spatial units including county subdivisions, counties, and metropolitan areas to accommodate diverse analytical applications.

## 2 Methods

### 2.1 Geographic coverage and social infrastructure Point of Interest

The objective of this data-descriptor study is to produce a new STU dataset that spans the 49 continental states and Washington D.C. in the United States. Based on the 2019 vintage geographic map that can be matched with our time use data, our study area covers 73,642 census tracts, 18,773 county subdivisions, 3,242 counties, and 939 metropolitan areas. Neighborhoods and communities at each spatial level were characterized by socio-demographic indicators from American Community Survey (ACS) 1-year estimation data published from 2019 to 2022[28,29]. We adjusted and aggregated neighborhood ACS indicators post-2020, originally aligned with 2020 geographic boundaries, to match 2019 vintage geographic boundaries using interpolation weights provided by NHGIS geographic crosswalks[30].

STU data were processed from visitation and time use records at social infrastructure POIs (i.e. Point Of Interest) from mobile devices[31]. A POI is a specific physical and non-residential location that includes a specific location and brand name[31]. For each POI, the record includes the North American Industry Classification System (NAICS) code developed by the U.S. Census Bureau, detailed to the 6-digit level, along with the street address, latitude and longitude, census geographic code as well as the open and close date. We used the POI data provided by Advan to retrieve the social infrastructure POIs. The base dataset was sourced through web crawling, public APIs, third-party licensing, and inferential modeling, followed by a rigorous de-duplication and merging process for data cleaning[32]. It has been continuously updated every month since January 2019[33].

We identified POIs relevant to social infrastructure places based on the six-digit NAICS codes corresponding to venues that support social and leisure activities, including shopping, physical exercise, dining, entertainment, social interaction, and religious participation[8,34,35]. These activities are defined according to the ATUS activities[34] or "Experience-oriented" activities in the study of the Global Human Day (GHD), which shares the same activity classification as ATUS[36]. ATUS provides national data on a wide range of daily activities and demographics annually with approximately 26,000 participants since 2003. Its detailed coding lexicons are instrumental in identifying activities that promote social connections. We examined all six-digit NAICS associated with social infrastructure places which are particularly related to ATUS-coded social and cultural activities[37] with the following criteria: (1) contributing to increased exposure to social environments; and (2) covering experience-oriented activities like meals, active recreation, and passive or interactive social activities.

In total, we have included seven types of activities. Each of them matches a list of social infrastructure-related NAICS sectors that identify specific sets of POIs (see Table 1). The detailed list of NAICS codes included in this study is available in the produced dataset repository.

Table 1. Social Infrastructure Activities, POI Sectors and Selection Rationale

| ATUS/GHD activity categories | Matched NAICS sectors for POI identification | Rationale |
|---|---|---|
| Grocery shopping (Grocery) | Grocery stores and food markets | Grocery shopping and food markets function as "third places" that provide opportunities for social interaction and exposure for visitors[38]. |
| Consumer goods purchases (Consume) | All retailers, except for food and beverage stores, including grocery stores and food markets. Other retails include different categories of POIs (e.g., clothing and clothing accessories, furniture and home furnishings) | Retail centers such as shopping plazas are places for interactional and recreational activities[39]. Beyond their traditional role as economic units, they also function as important sites of social infrastructure. |
| Participating in sports, exercise, and recreation (Physical exercises or sports) | Sports and recreation instruction, golf courses and country clubs, skiing facilities, marinas, fitness and recreational sports centers, bowling centers, all other amusement and recreation industries, | Amusement and recreation places, majorly sports facilities enable community interaction, well-being and social cohesion[40]. |
| Attending sporting or recreational events (Events) | Sports teams and clubs, Racetracks, and other spectator sports | Sporting and recreational venues host events that foster both physical, emotional, and social connections among participants and volunteers. Such events develop social capital, which can lead to increased civic engagement, prosocial behavior, community cohesion, and overall well-being[41]. |
| Eating and drinking (Dining) | Food service contractors, caterers, mobile food services, drinking places (alcoholic beverages), full-service restaurants, limited-service restaurants, cafeterias, grill buffets, and buffets, snack and nonalcoholic beverage bars. | Dining places often refer to as "*third places*," provide psychological benefits similar to urban parks. They serve as preferred spots for relaxation and social interaction which were found to enhance public mental health[42]. |
| Arts and entertainment other than sports (Arts) | Theater companies and dinner theaters, dance companies, musical groups and artists, other performing arts companies, museums, historical sites, zoos and botanical gardens, nature parks and other similar institutions, amusement and theme parks, amusement arcades, casinos (except casino hotels), other gambling industries. | Arts/entertainment places facilitate social interaction and strengthen community inclusiveness by providing shared cultural experiences that enhance social connectedness[43,44]. |
| Religious and spiritual activities (Religious) | Religious organizations | Religious organizations and places foster regular interactions among attendees, which in turn build strong social ties, increase community engagement, and promote closer relationships. These shared experiences contribute to social cohesion, emotional support, and a sense of belonging, making religious spaces important hubs for community connection and well-being[45]. |

## 2.2 Weekly visitation and time use patterns of social infrastructure places

The Advan Weekly Patterns data includes aggregated visitation data for each POI over days in a week in the U.S. since January 2019[46]. The data is aggregated and fully anonymized, without individual users' demographic information [33]. The data is collected when a user uses cell phone applications and all available sources to geolocate the phone such as cell tower signal or GPS services. According to the documentation of the data provider, the panel of the platform includes 32 million total devices across the U.S., which is about 15% of U.S. phone users[47]. Visitation to certain POI is determined by accounting for a ping within the boundary of a POI geometric polygon [48]. The duration of a visit is determined by the length of time a device's ping remains within the POI's polygon.

Additionally, for a specific POI, the home locations of visitors are only available at the census block group level to mitigate privacy concerns. The home locations were determined by analyzing six weeks of data during nighttime hours (between 6 pm and 7 am), requiring a sufficient amount of evidence (total data points and distinct days)[49]. When calculating census tract-level visitation and population records, we aggregated the measures of the associated census block group to the census tract level by extracting their first 11-digit FIPS code[33].

For each POI record, time-use information is captured as a distribution of visit durations across predefined intervals, reported under the attribute *"bucketed dwell time"*. This attribute provides visitation rates for the following time buckets: 0–4, 5–10, 11–20, 21–60, 61–120, 121–240, and >240 minutes. Notably, due to the pre-defined time buckets in the raw data, we are unable to distinguish or exclude work-related activities at certain social infrastructure places, particularly when the dwell time of visit exceeds 240 minutes (e.g. 8 hours) within the largest time bucket. However, this time bucket accounts for only about 10% of overall visitation frequency on average, varying by sector. Additionally, work-related social connections are recognized as contributing to employee well-being and are therefore considered relevant to the scope[50,51]. As such, this limitation is not expected to affect subsequent analyses. Future studies may define alternative thresholds for computing STU measurements.

When preprocessing the raw dwell time data for neighborhoods, we have identified and addressed two key data limitations. First, raw Advan data assigned the same records of visitation and time use to co-located POIs, which could inflate the estimated STU by several times if a large cluster of POIs shared the same polygon[33]. To avoid duplication, we evenly distributed neighborhood total STU to co-located POIs, following methods for similar data[48]. Second, the raw mobile device-based visitation data were derived from a dynamic panel of device users that continuously expands and evolves over time. Such changes in the panel could cause temporal and spatial variability in the sampling rate of the data[46,52] (specified in the Technical Validation section). To mitigate the impact of changing panel size on the STU calculation, we have normalized the visitation frequency by dividing neighborhood panel sizes, based on monthly updated panel summaries[33].

## 2.3 Analytical methods
### 2.3.1 Measuring neighborhood social infrastructure time use over weeks

The weekly total time ($T_{i,k}$) spent by residents of the neighbourhood ($i$) visiting specific type ($k$) of social infrastructure places is estimated by visitation-weighted aggregation of visit time ($T_j$) across all social infrastructure places of that type. For each social infrastructure place ($j$), the visit time ($T_j$) is derived based on a "*bucket dwell time*" pattern, which classifies visitation time into discrete time buckets. The total time of visits ($T_j$) for social infrastructure place ($j$) is the summation of the products of the median dwell time for each time bucket and the corresponding visitation frequency. $T_{i,k}$ is then computed by Eq.1:

$$T_{i,k} = \sum_j T_j \times \frac{V_{i,j}}{V_j} \quad j \in N_k \qquad (1)$$

here, $V_{i,j}$ represents the total visits from neighborhood $i$ to social infrastructure place $j$; $j$ is an element of the join of $N_k$ which includes all social infrastructure places of the activity type ($k$). Based on the calculated $T_{i,k}$, we developed two foundational measurements to quantify the two critical dimensions of STU: the length of time use and depth of activity engagement.

The first foundational measurement is neighbourhood Per User STU ($T_{user,i}$), which quantifies the weekly average time spent at all social infrastructure places of neighborhood $i$. It represents the overall aggregated length of engagement in a certain type of activity among residents in a neighbourhood. $T_{user,i}$ is computed as the sum of $T_{user,i,k}$, which denotes the time spent by residents in the neighborhood ($i$) on certain types of social infrastructure places classified by our activity categories ($k$) in Table 1. This relationship is expressed by Eq. 2:

$$T_{user,i} = \sum_k T_{user,i,k} = \sum_k \frac{T_{i,k}}{D_i} \quad k \in K \qquad (2)$$

where $K$ represents the set of activity categories (Table 1). Given that the phone-based visitation records have fluctuated with the changing number of devices ($D_i$) covered in each neighborhood $i$, the average STU of each device user is proxied by dividing $T_{i,k}$ by $D_i$ of the week.

The second foundational measurement is Per Visit STU ($T_{visit,i}$), which indicates the average dwelling time of each visitation at all social infrastructure places from the neighborhood ($i$). This measure builds on the STU quantification by adding a description of the depth of engagement in an average visit. $T_{visit,i}$ is the weighted summary of Per Visit STU for each type of social infrastructure place ($T_{visit,i,k}$). This relationship is expressed by Eq. 3:

$$T_{visit,i} = \sum_k T_{visit,i,k} \times \frac{\sum_j V_{i,j}}{V_i} = \frac{\sum_k T_{i,k}}{V_i} \quad k \in K, j \in N_k \qquad (3)$$

### 2.3.2 Quantifying weekly neighborhoods STU diversity with Shannon's Diversity Index

This study introduces novel analytical measures built upon the two foundational measures. We first measured the diversity of time use at social infrastructure places for each neighborhood with Shannon's diversity index, which was originally used to quantify species biodiversity within an ecosystem. This index is adapted to assess the variety and allocation of time spent at different sectors of social infrastructure places classified by the six-digit NAICS codes [7]. Specifically, the Shannon diversity of time use ($H_{t,i}$) is computed using Eq. 4.

$$\begin{cases} H_{t,i} = -\sum_{s=1}^{m}(P_{time,s,i} \times \ln P_{time,s,i}) \\ P_{time,s,i} = \dfrac{\sum_{j=1}^{n_s} T_j \times \dfrac{V_{i,j}}{V_j}}{\sum_{s=1}^{m}\sum_{j=1}^{n_s} T_j \times \dfrac{V_{i,j}}{V_j}} \end{cases} \qquad (4)$$

Here, $m$ denotes the total number of distinct social infrastructure sectors visited by residents of a given neighborhood ($i$). The variable $P_{time,s,i}$ represents the proportion of *total time spent* at $s$th sector relative to the *total time spent* at all $m$ sectors; $n_s$ indicates the total number of social infrastructure places within the sector $s$. The scale of diversity indicator is reversed by multiplying the total by −1, so the higher value of $H_{t,i}$ corresponds to greater diversity in both visitation patterns and time allocation across social infrastructure sectors.

A higher value of $H_{t,i}$ represents two dimensions of social infrastructure engagement: *richness*, indicating a greater variety of sectors visited, and *evenness*, capturing a more balanced distribution of visits and time spent across these sectors. This time-weighted visitation diversity metric allows us to capture both the breadth (richness) and depth (evenness) of social exposure during a typical week, offering nuanced insights into patterns of social infrastructure utilization within urban neighborhoods.

### 2.3.3 Measuring regional spatial inequality of neighborhood STU

To quantify levels of inequality overall STU across neighborhoods, we creatively computed a Gini index of social infrastructure time use ($G_{STU}$). Analogous to the traditional Gini coefficient used to measure income inequality[53]. $G_{STU}$ captures disparities in the distribution of STU across geographic regions, yet only applied to scales larger than census tract – such as county subdivisions, counties, and metropolitan areas. A higher $G_{STU}$ indicates a more unequal STU distribution, where a disproportionately small subset of the population accounts for a large share of total STU. $G_{STU}$ is computed as follows in Eq. 5.

$$G_{STU} = 1 - 2 \sum_{i=1}^{n} P_{pop,i} \left( Y_i \times \left( T_{user,i} - T_{user,i-1} \right) \right) \tag{5}$$

Here, $n$ indicates the total number of neighborhoods in a larger spatial unit (i.e., a county, a city, or a metropolitan); $P_{pop,i}$ is the proportion of the population in the neighborhood $i$ over the total population of the region; $T_{user,i}$ denotes the Per User STU in the $i$-th neighborhood; $Y_i$ represents the cumulative proportion of Per User STU up to and including the $i$-th neighborhood. The neighborhoods are ordered by increasing values of $T_{user,i}$ to ensure proper construction of the Lorenz curve underlying the Gini calculation.

### 2.3.4 Multi-scale aggregation of STU measures

Based on the neighborhood-level STU fundamental measures, we also produced social infrastructure time use measurements at county subdivisions (i.e., cities, townships, and villages), counties, and metropolitan areas. These measures enhance the adaptability of the data for spatial analysis across various scales, improving the compatibility with other datasets that use different spatial units for decision-making at different administrative levels. To aggregate the neighborhood-level STU measures ($T_{user,i,k}$, $T_{dwell,i,k}$, $H_{t,i}$) to a larger geographic unit $c$, the corresponding measures ($T_{user,c,k}$, $T_{dwell,c,k}$, $H_{t,c}$) are calculated using Eq.6, Eq.7, Eq.8, respectively.

$$T_{user,c,k} = \frac{\sum_{i=1}^{n_c} T_{user,i,k} \times D_i}{\sum_{i=1}^{C} D_i} \tag{6}$$

$$T_{dwell,c,k} = \frac{\sum_{i=1}^{n_c} T_{dwell,i,k} \times V_i}{\sum_{i=1}^{C} V_i} \tag{7}$$

$$H_{t,c} = \frac{\sum_{i=1}^{n_c} H_{t,i} \times D_i}{\sum_{i=1}^{C} D_i} \tag{8}$$

In these equations, $n_c$ represents the number of neighborhoods within the geographic unit $c$. $D_i$ represents the neighborhood device numbers. $V_i$ indicates the total number of visits from the neighborhood $i$ to social infrastructure places. Both $D_i$ and $V_i$ are used as weighting metrics when aggregating neighborhood-level measures to larger spatial scales.

## 3 Data Records

The data produced for this research is available through Open Science Framework[54], a publicly accessible dataset. Included in the repository are the Python codes developed to process and generate the national datasets from the raw Advan data. The key data files are stored in comma-separated values (.csv) format and organized into folders by corresponding to four geographic levels: neighborhood (census tract), county subdivision, county, and metropolitan area. Each folder contains weekly tabular datasets covering the period from January 2019 to May 2024. We also included a dataset of panel coverage rates at census tract, county and state levels that described data representation over months.

The datasets can be updated when the new source data becomes available along with web-based interactive maps. Table 2 offers a summary and descriptions of common attributes for tables at the four geographic levels. The data also records STU measures of specific subgroups of social infrastructure places in designated activity categories. At each geographic level, the tabular data begins by recording STU for all social infrastructure places (e.g., Per_User_STU_all). This is followed by detailed STU measures for each ATUS category, identified by the suffix "_sector" (e.g., Per_User_STU_Art).

Table 2. STU data attribute description and inclusion in data tables of different levels

| Attribute | Description | Data type |
| --- | --- | --- |
| GEOID | Unique identifier of the geographic units. | String |
| Timestamp | Starting date of a week (Monday) in yyyy-mm-dd format. | Datetime |
| Per_User_STU | Average total time use at social infrastructure places is divided by the total sampling population (proxied by the number of mobile devices in the panel). Available in the tabular data for county subdivisions, counties, and metropolitan areas. | Float |
| Per_Visit_STU | Average time spent on social infrastructure places for each visit. Available in the tabular data for census tract, county subdivisions, counties, and metropolitan areas. | Float |
| Diversity | Shannon diversity of social infrastructure place access for neighborhood calculated by time use. Available in the tabular data for county subdivisions, counties, and metropolitan areas. | Float |
| Gini | Gini index of STU. Available in the tabular data for county subdivisions, counties, and metropolitan areas. | Float |

### 3.1 Mapping time use measures across the Continental United States

In our proposed dataset, STU measures are linked to ACS geographic units, including 73,642 census tracts, 18,773 county subdivisions, 3,242 counties, and 939 metropolitan areas, and stored with a unique GeoID at each spatial unit (Table 2). Figures 1-4 illustrate samples of the geographic distribution of the four STU measures, based on their median values for 2023. We analyze nationwide spatial autocorrelation with Moran's I values. To demonstrate multi-scale data granularity, we mapped the median value as an example statistic of daily Per User STU across all social infrastructure place categories by counties (Fig. 1.a), county subdivisions (Fig.1.b), and neighborhood (Fig.1.c).

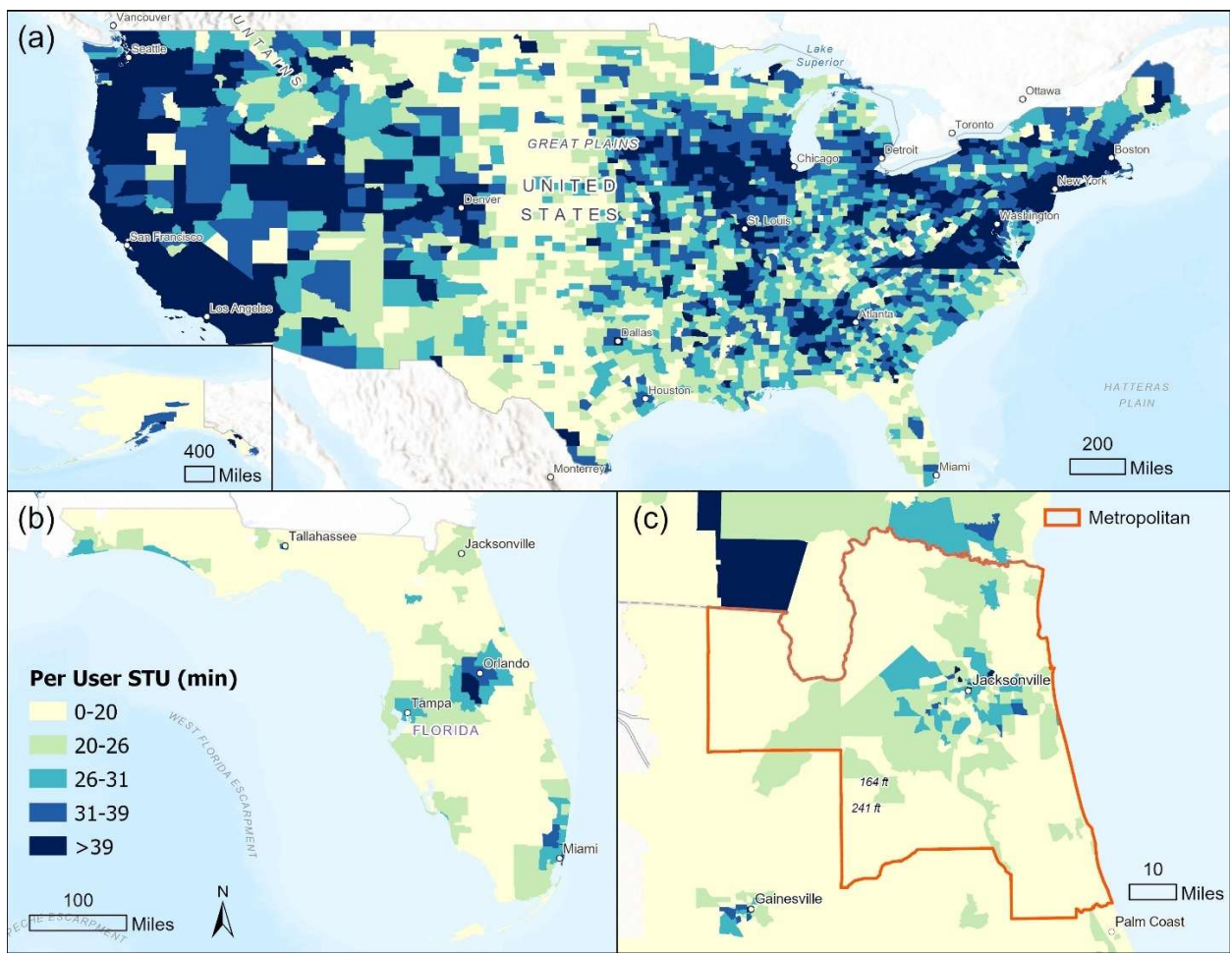

Fig. 1. Median daily Per User STU across the Continental United States in 2023. The map illustrate spatial variation in Per User STU at three geographic scales: (a) counties; (b) county subdivisions within the state of Florida; and (c) census tracts within the Jacksonville metropolitan area, FL. All maps uses a consistent legend with STU values grouped into five quantile-based categories and color-coded accordingly.

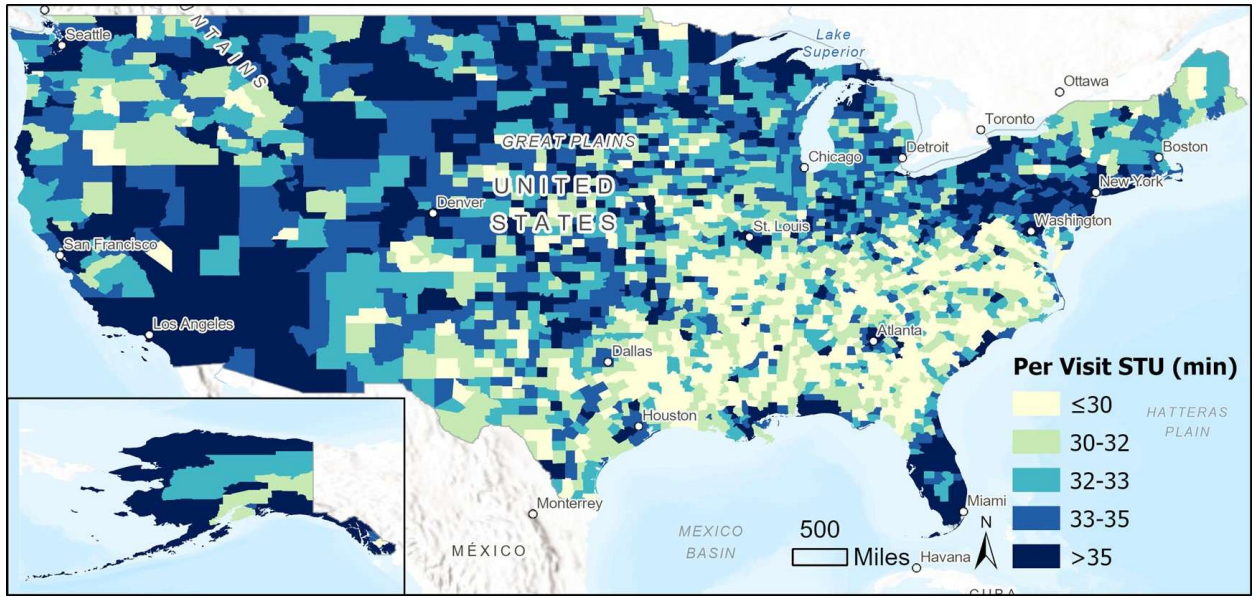

Fig. 2. Median Value of Per Visit STU for Counties Across the Continental United States in 2023: This map illustrates significant spatial autocorrelation of Per Visit STU measure. The Global Moran's I statistic is 0.256 (p-value: 0.000), reflecting a moderate level of clustering. The map's legend is color-coded into five quantile-based levels.

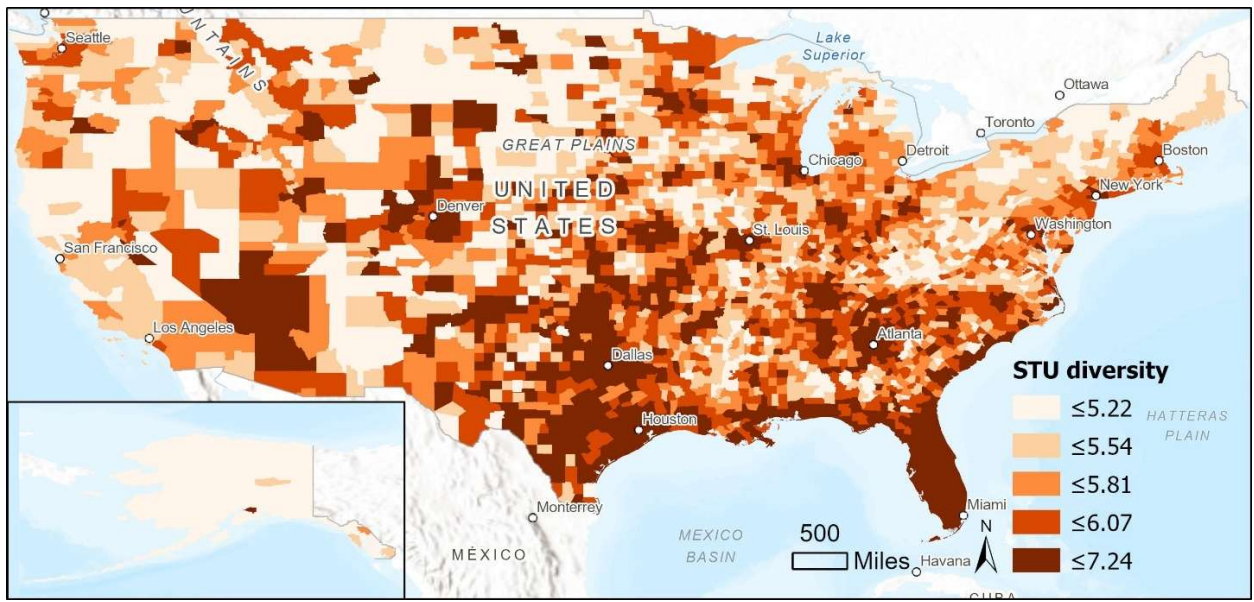

Fig. 3. Shannon Diversity Index of STU among neighborhoods across the Continental United States: This figure displays the spatial distribution of Shannon diversity in STU among neighborhoods, demonstrating significant spatial autocorrelation. The Global Moran's I statistic for this measure is 0.343 (p-value: 0.000), indicating notable clustering. The legend is color-coded into five levels based on quantiles.

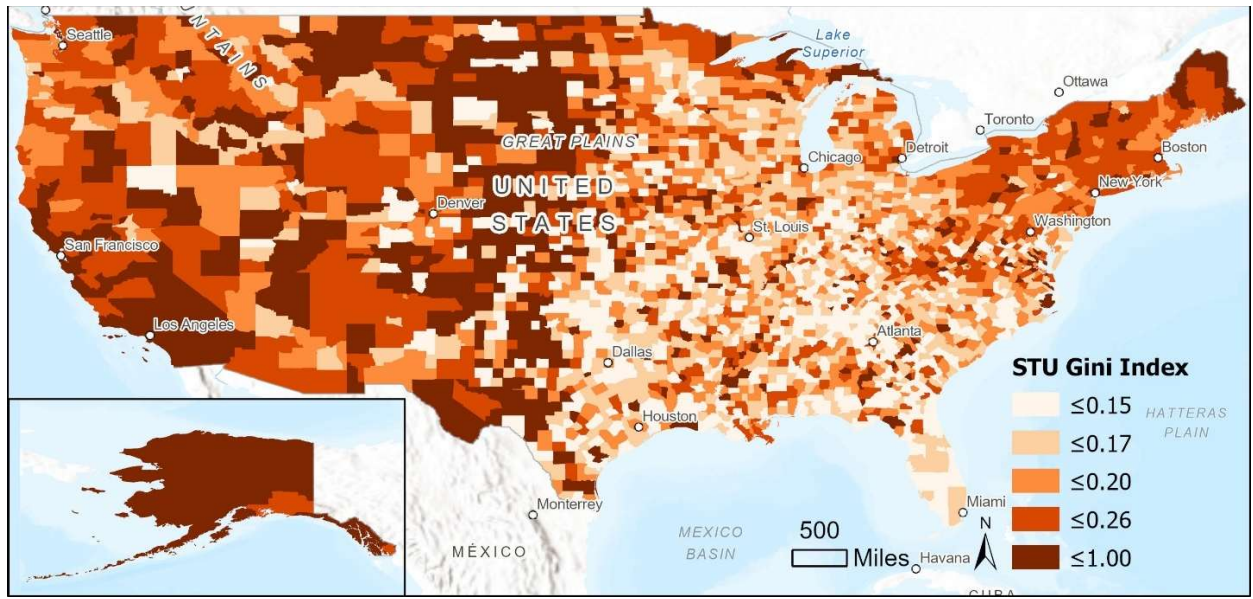

Fig. 4. Mapping of Gini Index of STU at the county level. This map visualizes the Gini index for STU, capturing inequality in time use across neighborhoods within each county. The Global Moran's I statistic is 0.169 (p-value: 0.000), indicating a relatively lower degree of spatial clustering. The legend is color-coded into five levels based on quantiles.

At the nationwide level, both foundational STU measures exhibit spatial clustering, with Per User STU (Global Moran's I: 0.359, p-value: 0.000) showing stronger clustering than Per Visit STU (Global Moran's I: 0.256, p-value: 0.000; Fig. 2). These patterns reflect clear regional disparities in STU across the U.S. Additionally, the nationwide geographic distribution of STU diversity (Fig. 3) and Gini coefficients (Fig. 4) show an inverse relationship: neighborhoods in the eastern half and northeastern regions of the U.S. exhibit higher spatial inequality and lower STU diversity, whereas neighborhoods in the southeastern regions show the opposite pattern. There appears to be a systematic pattern where access to a more diverse set of social infrastructure places can lead to more equitable STU.

### 3.2 Temporal change and distributions of foundational STU measures

Our datasets also enable the tracking of temporal changes in STU measures across different temporal units – weekly, monthly, and yearly. These temporal analyses enable further studies on how STU patterns evolve around disruptive events such as hurricanes and pandemics following infrastructure developments, or in response to new policies, employing time series statistics such as interrupted time series analysis [54].

Extant studies have examined the distribution patterns of human dynamics, such as mobility, often approximated by power law or log-normal distributions[55,56]. However, the distribution characteristics of STU remain largely unexplored. This gap limits efforts to benchmark local time use empirically and constraints modeling and simulation that integrate "time use" as a key variable for decision making. To address this, we analyzed the distributions of neighborhood-level Per User STU across social infrastructure places. We evaluated several candidate distributions – including normal, log-normal, Weibull, exponential, Gamma, power-law, Chi-squared – using Kolmogorov-Smirnov (KS) goodness-of-fit test. Among those, the log-normal distribution approximates the observed neighborhood STU with the best performance (see Table S1 and Table S2). Fig. 5 displays the observed distribution alongside fitted log-normal probability models for each year from 2019 to 2023. Notably, the scale parameter increased from 86.01 in 2019 to 271.2 in 2023, indicating a rising trend in social infrastructure time use across U.S. neighborhoods over the five-year period.

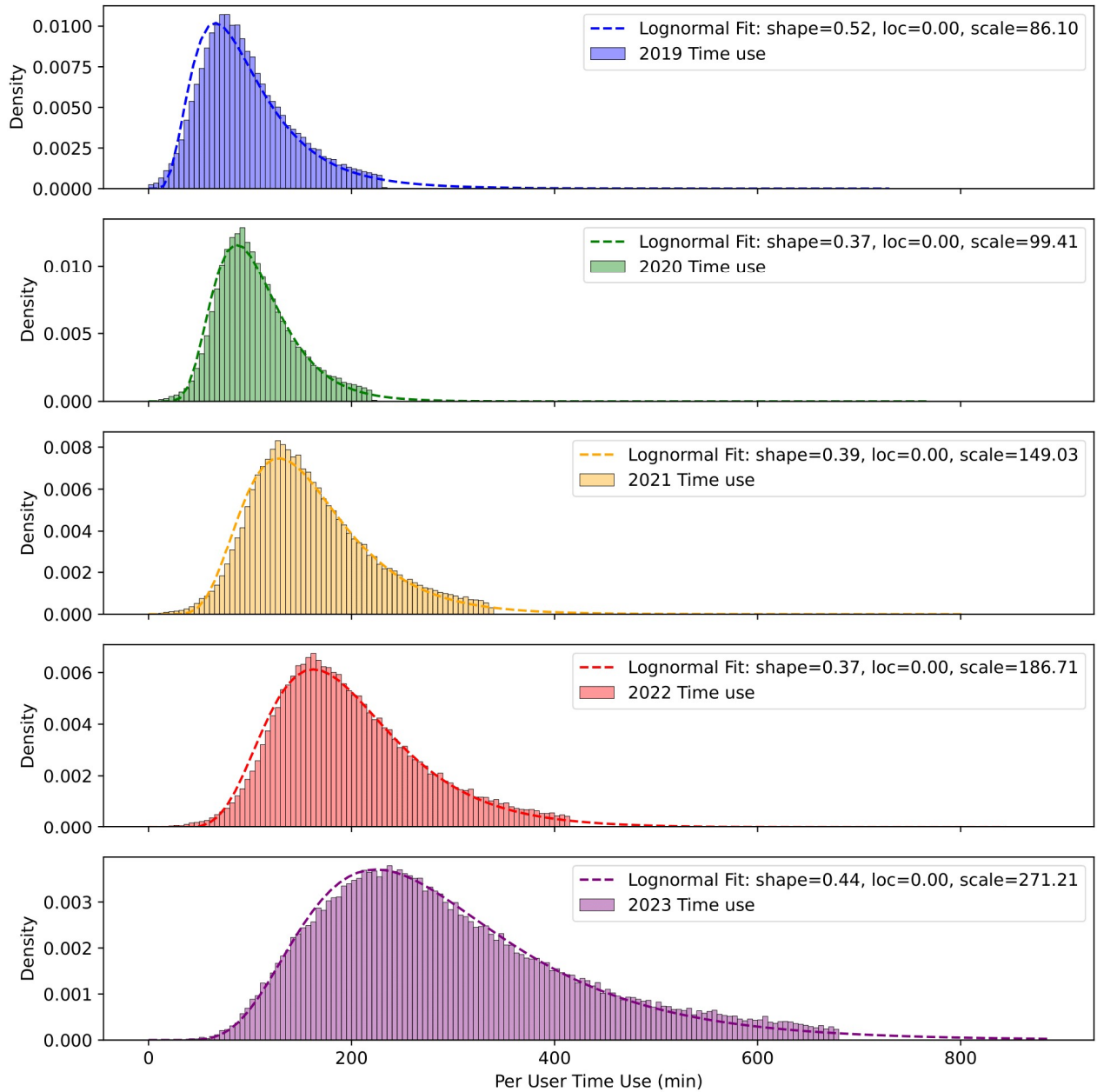

Fig. 5. Distributions of Per User STU from 2019 to 2023 with the fitted log-normal probability models. Each subplot shows the empirical distribution of neighborhood-level per-user STU alongside the best-fit log-normal probability model for a given year. KS statistics ranging from 0.169 to 0.280 indicate a moderate to strong fit. The shape parameter reflects skewness (i.e. a higher shape value indicates a more symmetric distribution centered around the mean, while a lower value implies a heavier skew toward shorter durations). The loc represents the minimum value of time use. The scale parameter captures the median value of the distribution, where a larger value correspond to greater time use.

Additionally, the fitted log-normal distributions of STU across activity categories vary in two dimensions (Table S2). First, differences in *scale parameters* reveal that three major activities–Art and Entertainment (Fig. 6.a), Eating and Drinking (Fig. 6.c), and Consuming Activities (Fig. 6.e)–exhibit higher median levels of neighborhood time use compared to other categories.  Second, variation in the *shape parameter* indicates differences in distributional skewness. Less routine and more event-driven activities, such as Art and Entertainment (shape = 5.98), Events (shape = 3.31), and Sports (shape = 1.79), show greater distributional skewness than more routine categories, which have shape values ranging from 0.44 to 0.67. This skewness reflects a highly uneven distribution of STU: most neighborhoods record low levels of time use, while a small subset exhibits substantially higher engagement. Such variation can serve as a proxy for understanding whether an activity is broadly accessed across neighborhoods and population groups or concentrated within select areas characterized by more intensive participation.

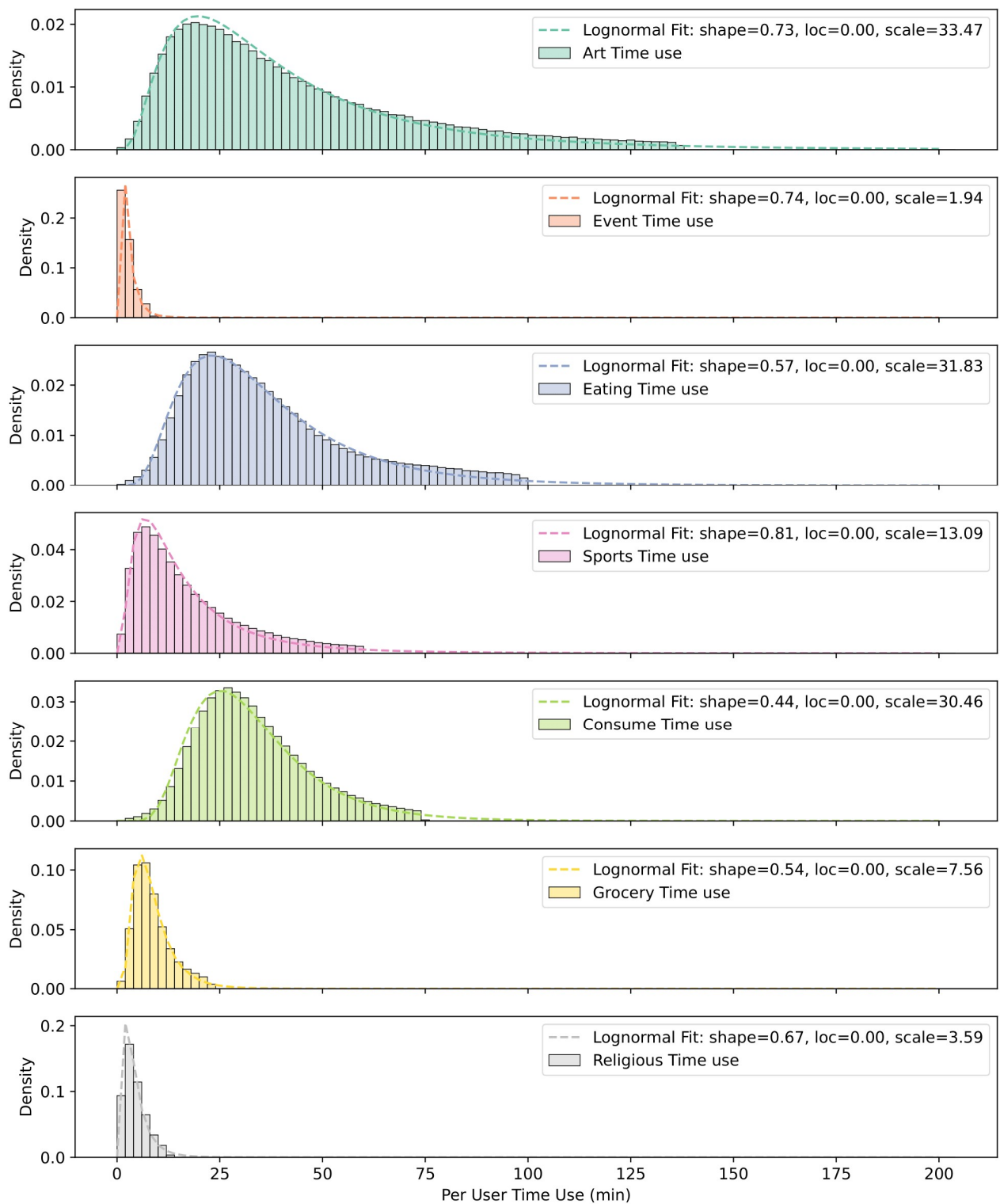

Fig. 6. Distributions of Per User STU across activity categories over 2019-2023 with fitted log-normal probability models. KS statistics from 0.156 to 0.350 indicate a moderate to strong fit of the model. The shape parameter denotes the skewness of the distribution. The loc represents the minimum STU value, while the scale determines the spread of the distribution.

## 4 Technical Validation

Our technical validation of the new datasets focuses on their population representativeness, alignment with activity-based time use reported in ATUS, and consistency with geographic and demographic patterns identified in prior empirical studies. These validation efforts are limited by the availability and resolution – both spatial and temporal – of suitable "ground-truth" datasets. Additionally, we offer guidance for users seeking to conduct rigorous analyses using these datasets. Future analyses can link our datasets with additional socio-economic, health, and environmental datasets for statistically rigorous implications of the STU measures. Our validation efforts focus on the two foundational measures: Per User STU and Per Visit STU.

### 4.1 Assessing sampling representativeness of panel size in raw data

Mobile device-based data offer unique opportunities for large-scale time use analysis but may be subject to spatial, temporal, and demographic biases. Our initial step examined how well the raw mobile phone user panel from Advan data represents local population size at census-tract, county, and state levels. The data reports of panel coverage rates – defined as the ratio of panel mobile devices to the overall population – are included in our dataset. Clear spatial variations have been revealed with notably lower representation in the western and northwest regions. These patterns are consistent with a prior study that validated the nation-level sampling rate[52]. Fig. 7 presents the panel coverage rates at the county level for 2019 when the national average rate was highest among all the years. To account for potential geographic representation issues in panel coverage rates, we normalized all STU measures to the dynamic panel size for each month. Since our targeted STU measures represent average rather than total values, changes in sampling rates primarily influence estimated time use in terms of their variability or uncertainty, rather than the accuracy of the mean values themselves.

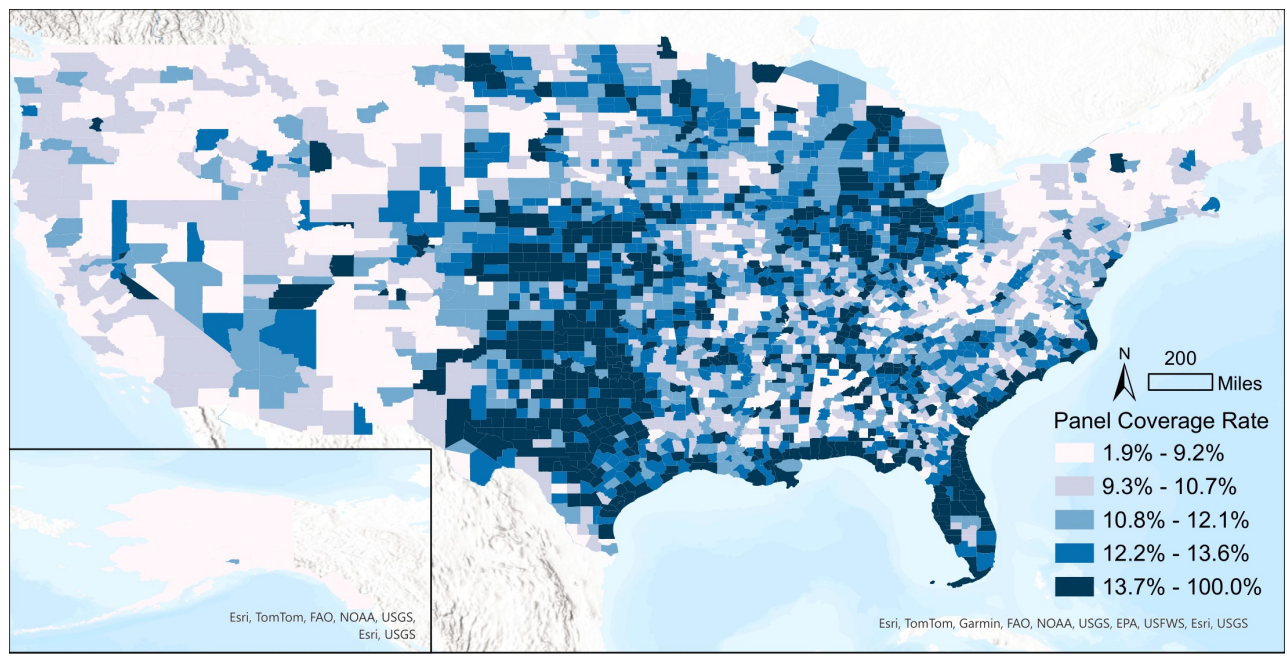

Fig. 7. Spatial distribution of panel coverage rate by census tract (percentage of population represented, 2019)

We further evaluated the alignment between the panel size and official population figures at multiple spatial scales. Using ACS data from 2019 to 2022, we calculated Pearson's correlations between monthly device counts and population size at census-tract, county and state levels (Fig. 8). Results show strong correlations at larger geographic scales, with over 90% alignment at the county and state levels. While correlations declined slightly following the onset of COVID-19, they have since stabilized. At the neighborhood (tract) level, the degree of alignment varied by year, ranging from approximately 54% to 72%. The year 2019 exhibited the most variability, likely due to being the first year of available data and should therefore be treated with particular caution in neighborhood-level analyses. To conclude, the panel size has robust representation at county and state levels, yet neighborhood-level panel size may not consistently correspond to local population size. Future data users need to note such data limitations and carefully interpret findings with reference to local representation rates provided in our dataset.

Time Profile of U.S. Neighborhoods: Datasets of Time Use at Social Infrastructure Places

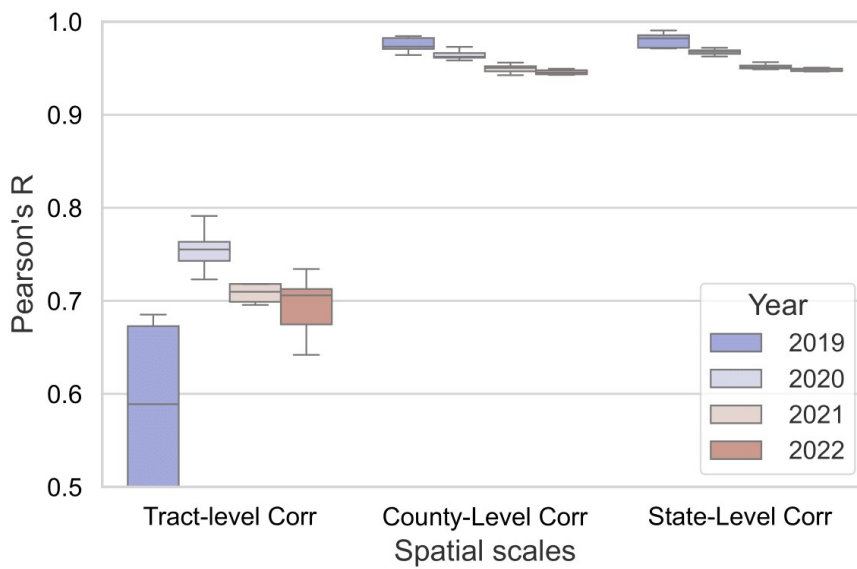

Fig. 8. Pearson's R Correlation Coefficient between panel mobile device numbers and local population across geographic scales and years

Although the aggregated Advan data limit direct examination of demographic bias, several factors suggest that its impact on our measures is minimal. First, data from the *National Health Interview Survey Wireless Substitution Report* [57] show that mobile phone adoption has increased steadily across the U.S. While 98% of Americans own a mobile phone in 2024, usage rates remain somewhat lower among individuals aged 50–64 (98%) and those 65 and older (94%)[58]. Second, recent evaluations [52,59] of SafeGraph mobile device panels (2018 – 2022) across demographic, and socioeconomic dimensions have found relatively minor biases among groups by gender, age, and moderate income, typically ranging from -0.05 to +0.05 (-0.05: 5% underrepresented in the data; +0.05: 5% overrepresented in the data). However, more pronounced underrepresentation was observed among Hispanic populations, low-income households, and individuals with lower educational attainment, with spatial and temporal variability depending on urbanicity and geographic levels.

Given the normalization of our STU metrics and the relatively small magnitude of identified biases, we anticipate a minor impact on the estimation of neighborhood-level STU measures. However, researchers using this dataset should carefully assess potential biases within the context of their specific study. We recommend evaluating representation bias across key demographic and socioeconomic groups, taking into account the selected temporal, spatial, and analytical scopes[60]. When bias is detected, it may be addressed through established methods such as post-stratification weighting, stratified analyses, or advanced techniques like propensity score matching or synthetic data augmentation to enhance the robustness of findings.

**4.2   Validation of foundational STU measures against ATUS data across social infrastructure activity types**

We conducted a comparative validation of both Per User STU and Per Visit STU across different activity types using national time use data from ATUS [12]. Given ATUS's regular updates and global relevance, the survey datasets serve as a benchmark for time-use research within the U.S.[36] While ATUS captures time spent both outside and at home, STU data captures only out-of-home time at social infrastructure places. Therefore, our validation focused on five activity categories that are primarily dependent on out-of-home visitation.

For the entire year of 2023, we validated the STU measures using the most recent dataset of ATUS, also surveyed in 2023, which represents a post-COVID-19 year. The range, mean, and standard deviation of weekly STU measures across U.S. neighborhoods are presented alongside the ATUS national summary for each corresponding measure. Specifically, we compared the ATUS metric "average time per day per civilian population" with "Per User STU," while the ATUS measure "average daily time for engaged individuals" was compared with "Per Visit STU."

Results show that ATUS-reported "average time per day per civilian population" falls within the range of Per User STU observed across U.S. neighborhoods, with activity duration rankings exhibiting general consistency between the two sources (Fig.9.a). An exception is STU value for "Eating and Drinking," which may underrepresent the total time use pattern for this activity type due to the exclusion of at-home activities not captured by STU data. Per Visit STU measures mirror the "average time per day for persons who engaged in the activity" in ATUS, though they are slightly lower consistently. This discrepancy likely arises because ATUS includes the total time spent on each activity throughout the day, whereas STU reflects only the duration of a single visit to a social infrastructure place for such activity.

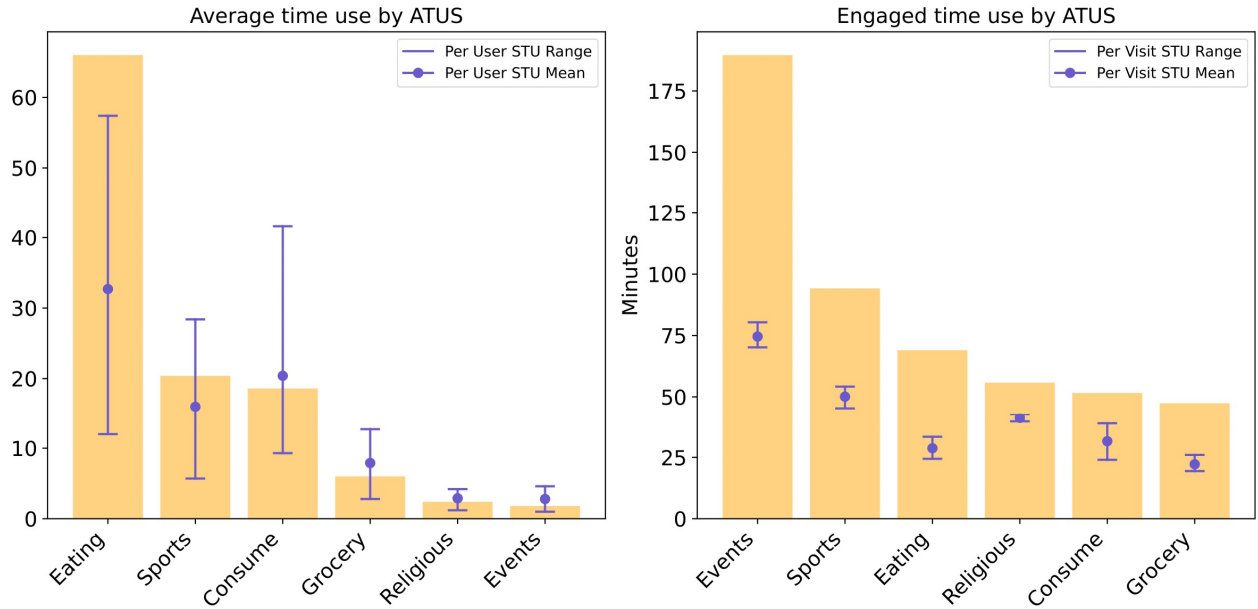

Fig. 9. Comparing ATUS average daily activity minutes with STU neighborhood daily (a) Per User STU (b) Per Visit STU

## 4.3   Aligned Yet More Nuanced Disparities in the new STU datasets, consistent with prior studies

Our validation builds on previous research that highlights disparities in activity engagement associated with geographic context and socio-demographic composition[12,61].

We first examined urban-rural disparities in time use patterns by comparing neighborhood-level STU distributions using Probability Density Functions (PDF), Cumulative Distribution Functions (CDF), and fitted lognormal distributions (Fig.10). The K-S test (K-S statistic: 0.531, P-value <0.001) confirms significant differences between the urban and rural distributions. Specifically, urban neighborhoods (distribution scale: 268.7) exhibit a higher proportion of areas with elevated Per User STU compared to rural neighborhoods (scale: 155.3). This pattern aligns with established findings on urban-rural disparity in physical activity participation levels and recreational accessibility[62,63], while offering a more spatially detailed and temporally continuous view enabled by the STU dataset.

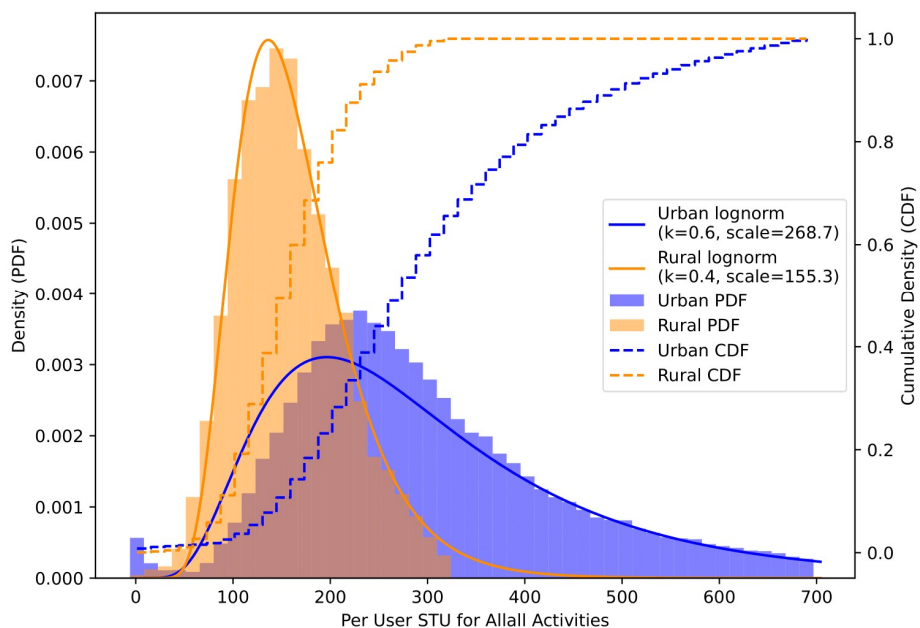

Fig. 10. Probability density and cumulative density distributions of Per User STU of urban and rural neighbourhoods

We further explored disparities in time use patterns among socio-demographic groups, building on prior studies that have used ATUS data to identify disparities in activity engagement across groups defined by race, gender, and educational attainment. To validate whether our datasets replicate these patterns at an aggregated neighbourhood level, we conducted *Pearson's R* correlation analysis between two foundational STU measures across various activity types and the demographic characteristics of U.S. neighbourhoods (Fig. 11).

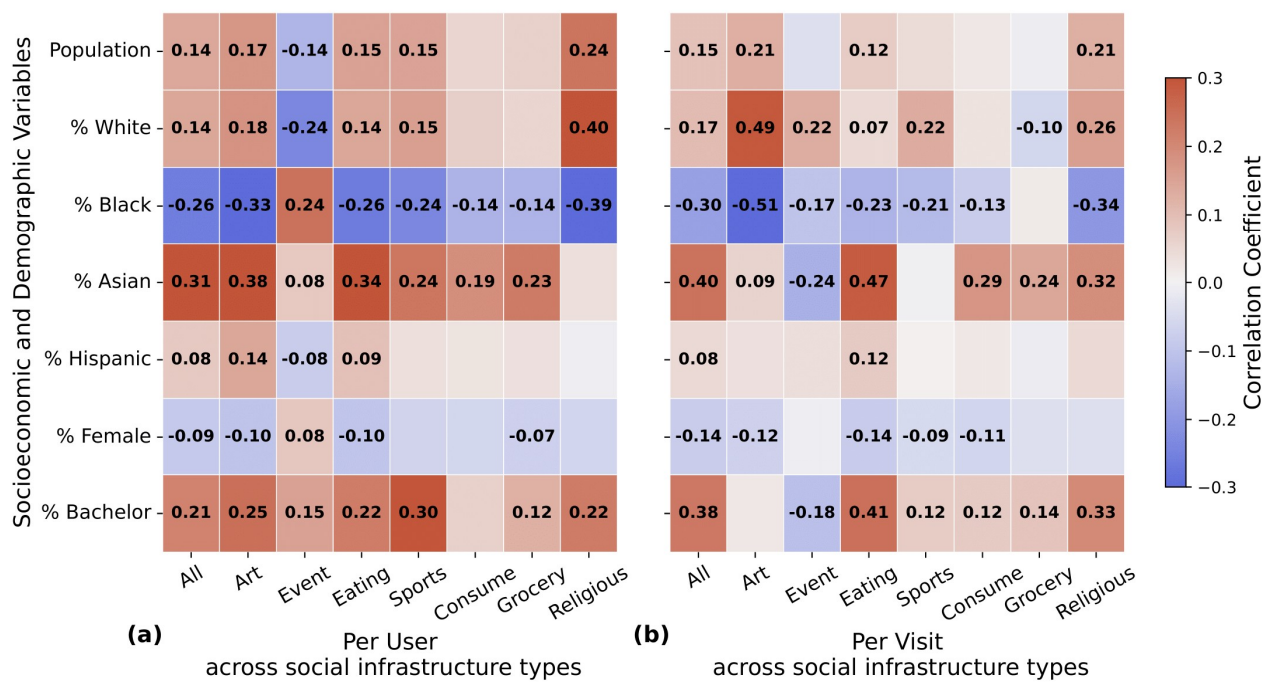

Fig. 11. Significant Pearson's R correlations ($p < 0.01$) between neighborhood socio-demographic patterns and two base STU measures across all and specific activities: (a) Per User STU; (b) Per Visit STU.

Our results are broadly consistent with findings from previous time use studies based on ATUS [61,64–66], supporting the validity and relevance of the new STU measures. For instance, a prior study has identified racial disparity has been observed: Black individuals typically spend less time in leisure activities, particularly in sports, than their White counterparts[64]. Our analysis supports this pattern at the neighborhood level: a higher percentage of the Black population in the neighbourhood is correlated with a lower total Per User STU (coefficient = -0.26, $p < 0.001$, 95% CI [-0.31, -0.21]) and total Per Visit STU (coefficient = -0.30, $p < 0.001$, 95% CI [-0.35, -0.25]), except for event-related STU. Conversely, neighbourhoods with a higher percentage of the White population are related with higher Per User STU (coefficient = 0.14, $p < 0.001$, 95% CI [0.09, 0.20]) and Per Visit STU (coefficient = 0.17, $p < 0.001$, 95% CI [0.12, 0.22]).

Gender-based inequities in leisure have also been consistently reported in ATUS research[66,67], often attributed to women's caregiving and housekeeping roles[65]. In line with these findings, our data show that a higher neighborhood percentage of females is significantly correlated with a lower total Per User STU (coefficient = -0.09, $p = 0.001$, 95% CI [-0.15, -0.04]). This negative correlation is particularly pronounced in activities such as visiting arts and entertainment venues (coefficient = -0.10) and eating and drinking places (coefficient = -0.10).

A study using ATUS data also found that individuals with higher educational levels tend to spend more time engaging in physical activities [61]. Consistent with this finding, we found that neighbourhoods with higher percentages of residents holding a bachelor's degree or higher are positively correlated with higher Per User STU at All social infrastructure places (Fig. 9; coefficient = 0.21, $p < 0.001$, 95% CI [0.15, 0.26]). This correlation was particularly strong for STU related to participation in sports, exercise, and recreation (coefficient = 0.30).

## Code Availability

The STU datasets for the 2019–2024 period are available through a repository published on Open Science Framework (https://doi.org/10.17605/OSF.IO/NSVJD). The repository also includes the code necessary to reproduce and validate the STU measures. An interactive ArcGIS Dashboard with map visualization is accessible through an online link (https://www.arcgis.com/apps/dashboards/95468aef197448c98283d1897fb9a37f).The raw Advan data was made available by Advan Research (https://advanresearch.com/) via subscriptions to Dewey Data platform (https://www.deweydata.io/). Validation datasets are publicly available: ACS socio-demographic data are extracted from https://data.census.gov/, geographic crosswalks are available at https://www.nhgis.org/geographic-crosswalks, and ATUS data can be accessed at https://www.bls.gov/tus/.

## Acknowledgments

This material is based on work supported by the National Science Foundation under Grant Nos. 2505675 and 2316450. Any opinions, findings, and conclusions or recommendations expressed in this material are those of the authors and do not necessarily reflect the views of the National Science Foundation.

## Author contributions

Y.W. conceptualized the research; Z.G. conducted the data processing and analysis; Y.W. and Z.G. worked on the results analysis and technical validation. Y.W. and Z.G. wrote the paper. Both authors edited and reviewed the manuscript.

## Competing interests

The authors declare no competing interests.